\documentclass[onecolumn, a4size, 12pt]{IEEEtran}
\usepackage{float}
\usepackage{amsmath,amsfonts,amssymb}
\usepackage{graphicx}
\usepackage{cite}
\usepackage{multirow}
\usepackage{algorithm}
\usepackage{algorithmic}
\usepackage{subfigure}
\usepackage{multirow}
\usepackage{url}
\usepackage{color}
\usepackage{array}
\usepackage{caption}

\graphicspath{{./figs/}}

\IEEEoverridecommandlockouts

\makeatletter
\newcommand*{\rom}[1]{\expandafter\@slowromancap\romannumeral #1@}
\makeatother

\renewcommand{\Re}{\mathrm{Re}}
\renewcommand{\Im}{\mathrm{Im}}

\newtheorem{theorem}{Theorem}

\newtheorem{lemma}{Lemma}
\newtheorem{corollary}{Corollary}

\newcommand{\xE}{\mathbb{E}}

\begin{document}
%\title{Energy Recycling: Simultaneous Energy Harvesting and Information Transmission for Wireless-Powered Communication}
\title{\LARGE Full-Duplex Wireless-Powered Relay with Self-Energy Recycling}
\author{\authorblockN{Yong~Zeng and Rui~Zhang\\} \vspace{-5ex}
\thanks{The authors are with the Department of Electrical and Computer Engineering, National University of Singapore.
(e-mail: \{elezeng, elezhang\}@nus.edu.sg).}
}
\maketitle

\begin{abstract}
This letter studies a wireless-powered amplify-and-forward relaying system, where an energy-constrained relay node assists the information transmission from the source to the destination using the  energy harvested from the source. We propose a novel two-phase protocol for efficient energy transfer and information relaying, in which the relay operates in full-duplex mode with simultaneous \emph{energy harvesting} and \emph{information transmission}.
Compared with the existing protocols, the proposed design possesses  two main advantages: i) it ensures uninterrupted information transmission since no time switching or power splitting is needed at the relay for energy harvesting; ii) it enables the so-called self-energy recycling, i.e., part of the energy (loop energy) that is used for information transmission by the relay can be harvested and reused in addition to the dedicated energy sent by the source. Under the multiple-input single-output (MISO) channel setup, the optimal  power allocation and beamforming design at the relay are derived. Numerical results show a significant throughput gain achieved  by our proposed design over the existing time switching-based relay protocol.
\end{abstract}

\section{Introduction}
 Radio-frequency (RF) enabled wireless energy transfer (WET) has recently emerged as a promising solution to provide convenient and perpetual power supply for energy-constrained networks. Since RF signal is able to convey both information and energy, one appealing direction of research is to jointly investigate  information and energy transfer to achieve simultaneous wireless information and power transfer (SWIPT) (see e.g. \cite{525,534} and references therein). An important application scenario for SWIPT lies in wireless-powered relaying (WPR), in which information is transmitted from the source to the destination via an energy-constrained relay node that is powered by means of WET. Existing studies on WPR  mostly consider half-duplex relaying and adopt either time switching-based relaying (TSR) \cite{539,541,535} or power splitting-based relaying (PSR) protocols \cite{535,536}, by exploiting the time-switching and power-splitting receiver architectures for SWIPT proposed in \cite{478}.
 In \cite{530} and \cite{537}, a time switching-based full-duplex wireless-powered transmission/relaying system is studied, where the source/relay node operates in full-duplex mode with simultaneous \emph{information reception  and energy/information transmission}.

 In this letter, we propose a new two-phase protocol for amplify-and-forward (AF) based WPR. In the first phase, information is transmitted from the source to the relay. In the second phase, the received signal at the relay is amplified and forwarded to the destination, and concurrently, dedicated energy signals are sent from the source to the relay for energy harvesting. Hence,  the relay operates in full-duplex mode in the second phase with simultaneous \emph{information transmission} and \emph{energy harvesting}. Compared with the existing TSR or PSR protocols, the proposed design possesses  the advantage of uninterrupted information transmission since no time switching or power splitting is needed at the relay  for energy harvesting. Besides, unlike the full-duplex relaying studied in \cite{537}, which suffers from severe self-interference and requires additional  energy consumption at the relay in order to implement the sophisticated analog and/or digital  interference-cancelation  \cite{540}, in our proposed full-duplex protocol, the self-interfering link at the relay is in fact beneficial since it enables the so-called self-energy recycling, i.e., part of the energy (loop energy) that is used for information transmission by the relay can be harvested and reused in addition to the dedicated energy sent by the source. Under the multiple-input single-output (MISO) channel setup, we study the optimal  power allocation and beamforming design at the relay to maximize the end-to-end throughput. We also study  the optimal time division ratio for the existing TSR as a benchmark for performance comparison. Numerical results show a significant throughput gain achieved by our proposed design.

\begin{figure}
\centering
\includegraphics[scale=0.7]{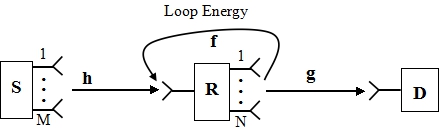}
\caption{Wireless-powered relaying with self-energy recycling.}\label{F:systemModel}
\end{figure}

\section{System Model and Proposed Protocol}
\begin{figure}
\centering
\includegraphics[scale=0.9]{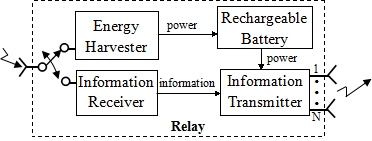}
\caption{A full-duplex wireless-powered relay with simultaneous energy harvesting and information transmission.} \vspace{-5pt} \label{F:fullDuplexRelay}
\end{figure}

As shown in Fig.~\ref{F:systemModel}, we consider a WPR system where the source node $\mathbf{S}$ transmits information to the destination node $\mathbf{D}$ via an energy-constrained relay $\mathbf{R}$, which is assumed to be solely powered by the energy harvested from $\mathbf S$.   We assume that $\mathbf{S}$ and $\mathbf{R}$ are equipped with $M$ and $N+1$  antennas, respectively, and $\mathbf{D}$ has one single  antenna. Furthermore, as illustrated in Fig.~\ref{F:fullDuplexRelay}, we assume that $\mathbf R$ %assigns one antenna for information/energy reception from $\mathbf S$ and the remaining $N$ antennas for information transmission to $\mathbf D$.
is equipped with $N$ RF chains for information transmission, as well as one RF chain  for information reception and one rectifier  for energy harvesting. %, which are both associated with the remaining antenna.
 Therefore, at each time instance, a maximum of $N$ antennas can be activated for information transmission and one antenna for information/energy reception.
For simplicity, we assume that the direct link from $\mathbf S$ to $\mathbf D$ is negligible and thus is ignored in this letter. In addition, we assume a quasi-static channel model with perfect channel state information (CSI) at $\mathbf S$ and $\mathbf R$. In practice, the CSI can be acquired by various methods, e.g., the pilot-assisted reverse-link channel training \cite{528}.

 %and one RF chain for information reception, which are connected to the $N$ transmitting antennas and one receiving antenna, respectively. One energy harvesting circuit is also equipped at the relay, which is connected

% which are associated with the $N$ transmitting antennas, as well as one information reception chain and one energy harvesting circuit, which are both connected to the receiving antenna via time switching.

%Amplify-and-forward (AF) relaying strategy is assumed due to its simplicity. In addition, we assume that the direct link between the source and destination nodes is negligible. % due to, e.g., a large separation distance and/or a highly adversary propagation environment in between the two nodes.
% We assume that the ET and ER/IT are equipped with $M$ and $N+1$  antennas, respectively, and the IR has one single antenna only.  All three nodes are assumed to be operated over the same frequency band. In addition, we assume that the direct link between the ET and IR is negligible due to, e.g., a large separation distance and/or a highly adversary propagation environment in between the two nodes.

\begin{figure}
\centering
\includegraphics[scale=0.6]{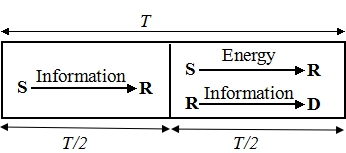}
\caption{Proposed protocol for full-duplex wireless-powered relaying.}\label{F:proposedProtocol}
\end{figure}

 %\section{Full-Duplex Relaying With Self-Energy Recycling}
 As shown in Fig.~\ref{F:proposedProtocol}, we propose a two-phase AF protocol for the WPR system. In the first phase of duration $T/2$, with $T$ denoting the total block duration, information is sent from $\mathbf S$ to $\mathbf R$ (with the switch shown in Fig.~\ref{F:fullDuplexRelay} connected to Information Receiver). In the second phase with the remaining time $T/2$, the received information signal at $\mathbf R$ is amplified and forwarded to $\mathbf D$ by its $N$ transmitting antennas, and concurrently, dedicated energy signals  are sent from $\mathbf S$ to the receiving antenna of $\mathbf R$ for energy harvesting (with the switch shown in Fig.~\ref{F:fullDuplexRelay} connected to Energy Harvester). The proposed protocol is elaborated in more details in the following section.

\section{Relaying with Self-Energy Recycling}
%In the first phase, information is transmitted from $\mathbb{S}$ to $\mathbb{R}$.
Denote by $\mathbf h\in \mathbb{C}^{M\times 1}$ the baseband equivalent MISO channel from $\mathbf S$ to the receiving antenna of $\mathbf R$. With the CSI $\mathbf h$ available at $\mathbf S$, the  maximal ratio transmission (MRT) with beamforming vector $\mathbf v_s=\mathbf h/\|\mathbf h\|$ is known to be optimal for information transmission. The information signal received at $\mathbf R$ during the first phase is thus given by
\begin{align}
y_{r,1}[k]=\sqrt{P_s} \|\mathbf h \| x_{s,1}[k]+ n_{r,1}[k], \label{eq:yr1}
\end{align}
where $k$ denotes the symbol index; $P_s$ represents the source transmission power; $x_{s,1}[k]$ denotes the information-bearing symbol sent by $\mathbf S$, which is assumed to be independently circularly-symmetric complex Gaussian (CSCG) distributed with zero mean and unit variance, i.e., $x_{s,1}[k]\sim \mathcal{CN}(0,1)$, $\forall k$; and $n_{r,1}[k]\sim \mathcal{CN}(0,\sigma_r^2)$ denotes the receiver AWGN noise at $\mathbf R$.

In the second phase, the received signal  in \eqref{eq:yr1} is amplified and forwarded to $\mathbf D$ by the $N$ transmitting antennas of $\mathbf R$ with power $P_r$ and beamforming vector $\mathbf v_r\in \mathbb{C}^{N\times 1}$, where $\|\mathbf v_r\|=1$. %The transmitted signal by the relay is then given by
 Denote by $\mathbf g\in \mathbb{C}^{N\times 1}$ the MISO channel from $\mathbf R$ to $\mathbf D$. The received signal at $\mathbf D$ can be expressed as
\begin{align}
y_d[k]&=\sqrt{P_r} \mathbf g^H \mathbf v_r \frac{y_{r,1}[k]}{\sqrt{A}} + n_d[k]\notag \\
&=\frac{\sqrt{P_sP_r}}{\sqrt{A}} \|\mathbf h\| \mathbf g^H \mathbf v_r x_{s,1}[k]+ \frac{\sqrt{P_r}}{\sqrt{A}} \mathbf g^H \mathbf v_r n_{r,1}[k] + n_d[k],\notag
\end{align}
where $A\triangleq P_s \|\mathbf h\|^2 + \sigma_r^2$ is the power of the received signal in \eqref{eq:yr1}; and $n_d[k]\sim \mathcal{CN}(0, \sigma_d^2)$ denotes the AWGN noise at $\mathbf D$. The end-to-end throughput from $\mathbf S$ to $\mathbf D$ can thus be expressed as $R=\frac{1}{2}\log_2(1+\gamma_d)$ in bps/Hz, with $\gamma_d$ denoting the received signal-to-noise ratio (SNR) at $\mathbf D$, which is given by
\begin{align}
\gamma_d = \frac{P_s \|\mathbf h\|^2}{\sigma_r^2 + \frac{\sigma_d^2 A}{P_r |\mathbf g^H \mathbf v_r|^2}}. \label{eq:gammad}
\end{align}

Concurrently with information transmission by the relay in the second phase, wireless energy is sent from $\mathbf S$ to the receiving antenna of $\mathbf R$ for energy harvesting. Particularly, the energy-harvesting circuitry at $\mathbf R$ not only harvests the dedicated energy sent from $\mathbf S$, but also recycles a portion of its own transmitted energy  via a loop channel denoted by $\mathbf f\in \mathbb{C}^{N\times 1}$. % the MISO channel of the loop link from the $N$ transmitting antennas of the  relay to its receiving antenna. %Due to the law of energy conservation, the loop power propagated via the loop link must be smaller than the transmitted power, i.e., $\|\mathbf f\|^2<1$ holds. We assume that energy and information beamforming are respectively performed at the ET and the IT, with beamforming vectors denoted as $\mathbf v_A$ and $\mathbf v_B$, respectively, where $\|\mathbf v_A\|=\|\mathbf v_B\|=1$. The equivalent baseband signal received at the ER can then be expressed as
The  received signal by the receiving antenna of $\mathbf R$ is expressed as
\begin{align}
y_{r,2}[k]=\sqrt{P_s} \|\mathbf h\| x_{s,2}[k] + \sqrt{P_r} \mathbf f^H \mathbf v_r \frac{y_{r,1}[k]}{\sqrt{A}} + n_{r,2}[k], \notag %\label{eq:yr2}
\end{align}
where $x_{s,2}[k]$ denotes the energy-bearing signal sent by $\mathbf S$, with $\xE[|x_{s,2}[k]|^2]=1$. %The signal in \eqref{eq:yr2} is then passed to the energy-harvesting circuitry for energy harvesting \cite{}.
By substituting $y_{r,1}[k]$ given in \eqref{eq:yr1} into the above equation and ignoring the negligible energy harvested from the receiver noise $n_{r,1}[k]$ and $n_{r,2}[k]$, the total harvested energy at  $\mathbf R$ during each block is given by
\begin{align}
E_r &= \frac{T}{2}\eta P_s \|\mathbf h\|^2 \xE\left[\Big| x_{s,2}[k]+ \frac{\sqrt{P_r}}{\sqrt{A}}\mathbf f^H \mathbf v_r x_{s,1}[k]\Big|^2\right]\\
&\leq \frac{T}{2}\eta P_s \|\mathbf h\|^2 \left(1+ \frac{\sqrt{P_r}}{\sqrt{A}}|\mathbf f^H \mathbf v_r| \right)^2, \label{eq:inequality}
\end{align}
where $0<\eta \leq 1$ denotes the energy harvesting efficiency at $\mathbf R$. The upper bound of harvested energy in \eqref{eq:inequality} is attained  when $x_{s,2}[k]=x_{s,1}[k] e^{j \angle \mathbf f^H \mathbf v_r}$, $\forall k$, with $\angle z$ denoting the phase of the complex number $z$. As will be shown in Theorem~\ref{theo:solP1}, with the optimal beamforming $\mathbf v_r^\star$, we have $\angle \mathbf f^H \mathbf v_r^\star=0$; hence, the above condition reduces to $x_{s,2}[k]=x_{s,1}[k]$, $\forall k$.
 Furthermore, to ensure that the average energy used for transmission by $\mathbf R$ does not exceed  that being harvested, we have $P_r T/2\leq E_r$. Let $\gamma_1$ and $\gamma_2$ be the SNRs of the first and second hops, respectively, i.e., $\gamma_1\triangleq P_s\|\mathbf h\|^2/\sigma_r^2$, and $\gamma_2\triangleq P_r |\mathbf g^H \mathbf v_r|^2/\sigma_d^2$.
Using \eqref{eq:gammad} and \eqref{eq:inequality} and by discarding the constant terms, the  problem of maximizing $R$ is equivalent to that of maximizing $\gamma_2$, which can be formulated as
\begin{align}
\mathrm{(P1)}: & \ \underset{P_r,\mathbf v_r}{\max}\   \gamma_2\triangleq P_r |\mathbf g^H \mathbf v_r|^2/\sigma_d^2 \notag \\
\text{ s.t. } & 0\leq P_r \leq \eta P_s \|\mathbf h\|^2 \left(1+ \frac{\sqrt{P_r}}{\sqrt{A}}|\mathbf f^H \mathbf v_r| \right)^2,\\
& \|\mathbf v_r\|=1.
\end{align}
%(P1) is non-convex in general. However, after some manipulations,
 Let $0\leq \theta\leq \pi/2$  be the effective angle between $\mathbf f$ and $\mathbf g$, which is defined as %\footnote{Note that this definition slightly differs from the standard definition: $\cos(\theta)\triangleq \Re(\mathbf f^H \mathbf g)/(\|\mathbf f\| \|\mathbf g\|)$, for which $\theta$ varies from $0$ to $\pi$.}
%\begin{align}
$\cos(\theta)\triangleq |\mathbf f^H \mathbf g|/(\|\mathbf f\| \|\mathbf g\|)$.
%\end{align}
%We have the following result:
%The optimal solution to (P1) is then given by the following theorem.
We then have
\begin{theorem}\label{theo:solP1}
The optimal solution $(P_r^\star, \mathbf v_r^\star)$ to (P1) is
\begin{align}
P_r^\star= \|\mathbf v^\star\|^2, \ \mathbf v_r^\star=\mathbf v^\star /\|\mathbf v^\star\|,
\end{align}
where
%\begin{align}
$\mathbf v^\star= \alpha_1 e^{j \angle \mathbf g^H \mathbf f} \mathbf g+ \alpha_2 \mathbf f$,
%\end{align}
with
\begin{align}
\alpha_1&=\frac{\|\mathbf h\|\sqrt{\left(1+\frac{1}{\gamma_1}\right) \eta P_s}}{\|\mathbf g\| \sqrt{1+\frac{1}{\gamma_1} -\eta \|\mathbf f\|^2 \sin^2 \theta}}, \text{ and} \notag \\
\alpha_2 &=  \frac{\eta \|\mathbf h\| \sqrt{\left( 1+\frac{1}{\gamma_1}\right)P_s}}{1+\frac{1}{\gamma_1}-\eta \|\mathbf f\|^2}\left(1+ \frac{\sqrt{\eta} \|\mathbf f\| \cos \theta}{\sqrt{1+\frac{1}{\gamma_1}-\eta \|\mathbf f\|^2 \sin^2 \theta}} \right).\notag
\end{align}
The corresponding optimal value of (P1) is %\eqref{eq:vstar} shown at the top of the next page,
%\begin{figure*}
\begin{equation}\label{eq:vstar}
%\small
\begin{aligned}
%\gamma_2^\star=\frac{\left(1+\frac{1}{\gamma_1}\right)\eta P_s \|\mathbf h\|^2 \|\mathbf g\|^2}{\sigma_d^2\left(1+\frac{1}{\gamma_1}-\eta \|\mathbf f\|^2\right)^2}
%\left(\sqrt{\eta}\|\mathbf f\| \cos\theta + \sqrt{1+\frac{1}{\gamma_1}-\eta \|\mathbf f\|^2 + \eta \|\mathbf f\|^2 \cos^2\theta} \right)^2
\gamma_2^\star= & \frac{\left(1+\frac{1}{\gamma_1}\right)\eta P_s \|\mathbf h\|^2 \|\mathbf g\|^2}{\sigma_d^2\left(1+\frac{1}{\gamma_1}-\eta \|\mathbf f\|^2\right)^2}
\left(\sqrt{\eta}\|\mathbf f\| \cos\theta + \sqrt{1+\frac{1}{\gamma_1}-\eta \|\mathbf f\|^2 \sin^2\theta } \right)^2,
\end{aligned}
\end{equation}
%\end{figure*}
and the maximum  throughput from $\mathbf S$ to $\mathbf D$ is
%\begin{align}
$R^\star=\frac{1}{2}\log_2\left(1+\frac{\gamma_1 \gamma_2^\star}{\gamma_1+\gamma_2^\star +1}\right)$.
%\end{align}
\end{theorem}
\begin{IEEEproof}
Please refer to Appendix~\ref{A:solP1}.
\end{IEEEproof}
Theorem~\ref{theo:solP1} indicates  that the optimal relay  beamforming is obtained as  a linear combination of $\mathbf g$ and $\mathbf f$ so as to achieve a good balance between information transmission and self-energy recycling. Furthermore, \eqref{eq:vstar} shows that $\gamma_2^\star$ monotonically increases with $\cos(\theta)$, i.e., as the MISO channels $\mathbf f$ and $\mathbf g$ are better aligned.
%Note that throughput the paper, we assume that perfect channel sate information (CSI) is available at each node. %The problem formulations for these two scenarios are given next.

%Next, we examine the optimal solution to (P1) for the special case where the relay has only one transmitting antenna, i.e., $N=1$, and thus the MISO channels $\mathbf f$ and $\mathbf g$ reduce to two complex scalars $f$ and $g$, respectively. In this case, no beamforming can be applied at the relay, and the optimal transmission power can be obtained by directly applying Theorem~\ref{theo:solP1} by noting that $\cos(\theta)=1$, which gives the following corollary.

\begin{corollary}\label{cor:SISOP1}
In the special case when $\mathbf R$ has one single  transmitting antenna ($N=1$) and the MISO channel $\mathbf f$ becomes a  complex scalar $f$,   the solution to (P1) reduces to
\begin{align}
P_{r}^{\star}=\frac{\eta P_s \|\mathbf h\|^2}{\left( 1-\sqrt{\eta}|f|/\sqrt{1+\frac{1}{\gamma_1}}\right)^2}.\label{PrstarSISO}
\end{align}
%and the corresponding optimal value is
%\begin{align}
%\gamma_2^{\star}=\frac{\left( 1+ \frac{1}{\gamma_1}\right)\eta P_s \|\mathbf h\|^2 |g|^2 }{\sigma_d^2\left(\sqrt{1+\frac{1}{\gamma_1}}-\sqrt{\eta} |f|\right)^2}. \label{eq:R1starSISO}
%\end{align}
\end{corollary}
It follows from \eqref{PrstarSISO} that $P_r^\star > \eta P_s \|\mathbf h\|^2$, i.e., thanks to the self-energy recycling in the proposed protocol, the relay can transmit with larger power  than that using the directly harvested energy from the source.

\begin{figure}
\centering
\includegraphics[scale=0.65]{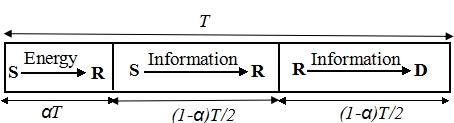}
\caption{Time-switching based wireless-powered relaying \cite{535}.}\label{F:TSRelaying}
\end{figure}

\section{Time-Switching Based Relaying}
In this section, we consider  the existing TSR protocol \cite{535} for our studied WPR system, which serves as a benchmark for performance comparison. As shown in Fig.~\ref{F:TSRelaying}, the TSR protocol in general consists of three phases. In the first phase of duration $\alpha T$, with $0<\alpha <1$, energy signal is sent from $\mathbf S$ to $\mathbf R$ for energy harvesting. In the second and third phases each with equal duration $(1-\alpha)T/2$, information is sent from $\mathbf S$ to $\mathbf R$ and then forwarded from $\mathbf R$ to $\mathbf D$ via AF relaying, respectively.
%With source transmission power $P_s$ and MRT beamforming $\mathbf v_s=\mathbf h/\|\mathbf h\|$, the total energy harvested during each block can be expressed as
%\begin{align}
With this protocol, the total energy harvested by $\mathbf R$ during each block is $E_r=\alpha T \eta P_s \|\mathbf h\|^2$, and the transmit power of $\mathbf R$ can be obtained as $P_r=E_r/((1-\alpha)T/2)=2\alpha  \eta P_s \|\mathbf h\|^2/(1-\alpha)$. Using \eqref{eq:gammad} as well as the fact that the optimal transmit  beamforming at $\mathbf R$ should match to the MISO channel $\mathbf g$ in this case, i.e., $\mathbf v_r=\mathbf g/\|\mathbf g\|$, the received SNR at $\mathbf D$ can be expressed as
\begin{align}
\gamma_d=\frac{\gamma_1}{1+ \frac{C(1-\alpha)}{\alpha}},
\end{align}
where  $C\triangleq (1+\gamma_1)\sigma_d^2/(2 \eta P_s \|\mathbf h\|^2\|\mathbf g\|^2)$. %The corresponding throughput in bps/Hz is
%\begin{align}
%R(\alpha)=\frac{1-\alpha}{2} \log\left(1+ \frac{\gamma_1}{1+ \frac{C(1-\alpha)}{\alpha}} \right), \label{eq:Ralpha}
%\end{align}
 The throughput maximization problem in this case is then formulated as
\begin{align}
\mathrm{(P2)}: & \ \underset{0\leq \alpha \leq 1}{\max}\  R(\alpha)\triangleq  \frac{1-\alpha}{2} \log\left(1+ \frac{\gamma_1}{1+ \frac{C(1-\alpha)}{\alpha}} \right). \notag
\end{align}
\begin{theorem}\label{theo:solP2}
The optimal solution to (P2) is
\begin{align}
\alpha^\star= \frac{(z^\star-1)C}{(z^\star-1)C+1+\gamma_1-z^\star},
\end{align}
where $1<z^\star <1+\gamma_1$ is the unique solution of the equation $f(z)=0$ in the interval $(1,1+\gamma_1)$, with $f(z)$ given by
\begin{align}
f(z)\triangleq &\gamma_1 C z \ln z + (C-1) z^2 -z (\gamma_1 C + 2C -2 \gamma_1 -2)\notag \\
& -(\gamma_1+1)(\gamma_1+1-C),\label{eq:fz}
\end{align}
which is a monotonically increasing function over $z\in(1,1+\gamma_1)$ with $f(1)=-\gamma_1^2$ and $f(1+\gamma_1)=\gamma_1 C(1+\gamma_1)\ln(1+\gamma_1)$.
\end{theorem}
\begin{IEEEproof}
Please refer to Appendix~\ref{A:solP2}.
\end{IEEEproof}
%With Theorem~\ref{theo:solP2}, the optimal time sharing $\alpha^\star$ for energy transmission can be efficiently obtained with bisection method based on \eqref{eq:fz}.

\begin{figure}
\centering
\includegraphics[scale=0.25]{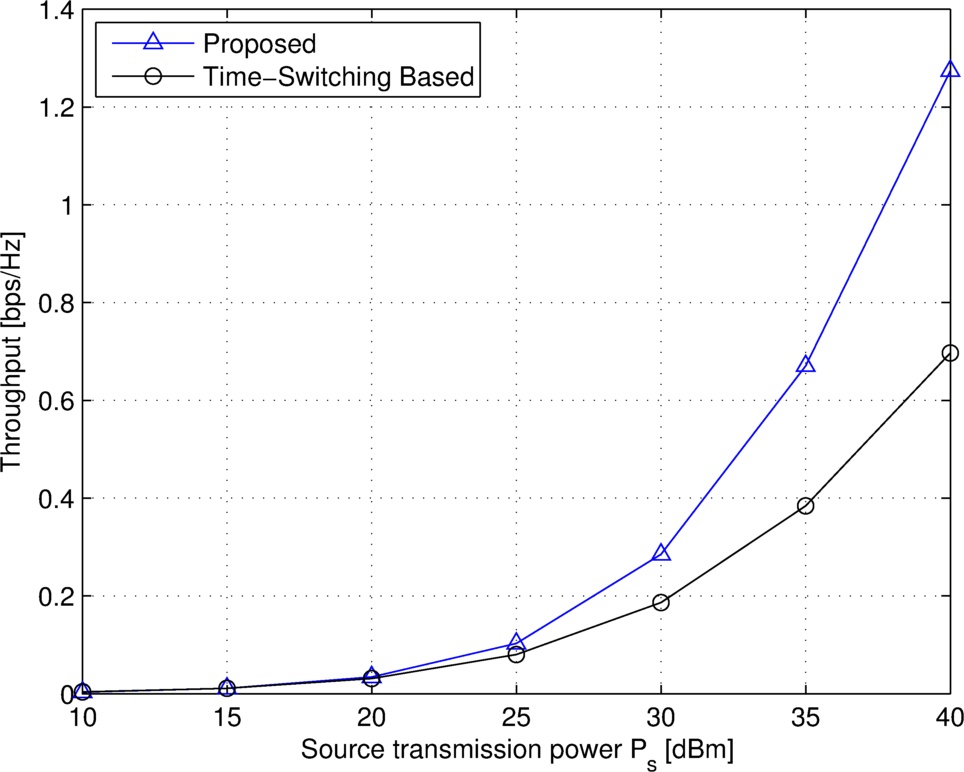}
\caption{Throughput versus source transmission power $P_s$.}\label{F:ThroughputvsPs}
\end{figure}

\section{Numerical Results}
%In this section, numerical results are provided to corroborate our study.
 We assume that the available bandwidth for the WPR system is $10$MHz, and the AWGN power spectrum density at both the receivers of $\mathbf R$ and $\mathbf D$ is $-160$dBm/Hz. Hence, the total noise power is $\sigma_r^2=\sigma_d^2=-90$dBm. The energy harvesting efficiency at $\mathbf R$ is assumed to be $\eta = 0.8$. We also  assume that the transmitting antennas at both $\mathbf S$ and $\mathbf R$ form a 2-element ($M=N=2$) uniform linear array (ULA) with elements separated by distance $d=\lambda/2$, where $\lambda$ denotes the carrier wavelength. Furthermore, we assume that the receiving antenna at $\mathbf R$ has the same distance $\lambda/2$ to its two transmitting antennas, so that the loop channel can be modelled as $\mathbf f=\sqrt{\beta_{rr}}[\begin{matrix}1 & 1\end{matrix}]^T$, where $\beta_{rr}=-15$dB denotes the loop-link path loss \cite{542}. We assume that both $\mathbf h$ and $\mathbf g$ are modelled  by line-of-sight (LoS) channels with angle-of-departure (AoD) $\theta_h=10^\circ$ and $\theta_g=5^\circ$, respectively. Thus, we have $\mathbf h=\sqrt{\beta_{sr}}[1, e^{j2\pi d/\lambda\sin \theta_h}]^T$ and $\mathbf g=\sqrt{\beta_{rd}}[1, e^{j2\pi d/\lambda \sin \theta_g}]^T$, with $\beta_{sr}$ and $\beta_{rd}$ denoting the path loss from $\mathbf S$ to $\mathbf R$ and from $\mathbf R$ to $\mathbf D$, respectively. We set $\beta_{sr}=\beta_{rd}=-60$dB, which may correspond to a separation distance around $30$ meters with carrier frequency at $900$ MHz.%  $reflects the fact that the propagation distance of the loop link is much shorter than the other two links.

In Fig.~\ref{F:ThroughputvsPs}, the throughput in bps/Hz is plotted against the source transmission power $P_s$ for both the  proposed design and the TSR protocol \cite{535} with optimal time allocation  $\alpha^\star$. It is observed  that significant throughput gains are achieved by our proposed protocol with sufficiently large $P_s$,  thanks to the self-energy recycling as well as the simultaneous energy harvesting and information transmission at the relay, which ensures continuous information transmission without any interruption due to energy transfer.

\section{Conclusion}
In this letter, we propose  a new protocol for wireless-powered relaying system with simultaneous energy harvesting and information transmission at the relay, which achieves  uninterrupted information transmission and also self-energy recycling. The optimal power allocation and beamforming design at the relay are obtained under the MISO channel setup and assuming AF relaying. Numerical results show significant throughput gains with the proposed design over the existing scheme.
%especially with relatively large source transmission power $P_s$,

% antennas  receiving antenna of the relay     the source and relay nodes are equipped with a $2$-element uniform linear array (i.e., $M=N=2$) as transmitting antennas.

%We assume that $M=N=2$,

\appendices

\section{Proof of Theorem~\ref{theo:solP1}}\label{A:solP1}
By discarding the term $\sigma_d^2$, and defining $\mathbf v\triangleq \sqrt{P_r} \mathbf v_r$, $a\triangleq \eta P_s \|\mathbf h\|^2$, and $\hat{\mathbf f}=\mathbf f/\sqrt{A}$, (P1) can be recast as
\begin{align}
\mathrm{(P1.1)}: \ \underset{\mathbf v}{\max}\ & |\mathbf g^H \mathbf v|^2  \quad
\text{ s.t. }  \|\mathbf v\|^2 \leq a \Big( 1 + \big|\hat {\mathbf f}^H \mathbf v\big| \Big)^2.\notag
\end{align}

%(P1.1) is still non-convex since the objective function is non-concave and the constraint is non-convex. %In the following, we show that (P2.1)
\begin{lemma}\label{lemma:equivalance}
 (P1.1) is equivalent to the following problem
\begin{align}
\mathrm{(P1.2)}: \ &  \underset{\mathbf v}{\max}\ |\mathbf g^H \mathbf v|^2 \notag \\
\text{ s.t. } & \| \mathbf v\|^2 \leq a \Big( 1 + |\hat {\mathbf f}^H \mathbf v|^2+ 2 \Re (\hat{\mathbf f}^H \mathbf v) \Big). \label{eq:P1.4Constr}
\end{align}
\end{lemma}
\begin{IEEEproof}
First, it is observed that if $\mathbf v$ is an optimal solution to (P1.1), so does $\mathbf v e^{j\omega}$ for arbitrary phase rotation $\omega$. Thus, without loss of generality, we can restrict that $\hat{\mathbf f}^H \mathbf v$ in (P1.1) is a real number. As a result, (P1.1) is equivalent to
\begin{equation}\label{eq:imeq0}
\begin{aligned}
 \quad \underset{\mathbf v}{\max}\ & |\mathbf g^H \mathbf v|^2  \\
\text{ s.t. } & \|\mathbf v\|^2 \leq a \Big( 1 + \Re( \hat{\mathbf f}^H \mathbf v ) \Big)^2,\\
& \Im (\hat{\mathbf f}^H \mathbf v)=0.
\end{aligned}
\end{equation}

Next, we prove that the last constraint of problem \eqref{eq:imeq0} is redundant by showing that the optimal solution $\mathbf v^\star$ to the relaxed problem without this constraint always satisfy $\Im (\hat{\mathbf f}^H \mathbf v^\star)=0$.
%\begin{equation}\label{P:relaxed}
%\begin{aligned}
% \quad \underset{\mathbf v}{\max}\ & |\mathbf g^H \mathbf v|^2 \\
%\text{ s.t. } & \|\mathbf v\|^2 \leq a \Big( 1 + \Re( \hat{\mathbf f}^H \mathbf v ) \Big)^2.
%\end{aligned}
%\end{equation}
Suppose, on the contrary, that $\Im (\hat{\mathbf f}^H \mathbf v^\star)\neq 0$. Let $\phi=\angle \hat{\mathbf f}^H \mathbf v^{\star}$. We then have
\begin{align}
\hspace{-4ex}\|\mathbf v^{\star}\|^2\leq a \Big( 1 + \Re(\hat{ \mathbf f}^H \mathbf v^{\star}) \Big)^2 < a \Big( 1 + \Re( \hat{\mathbf f}^H \mathbf v^{\star}e^{-j\phi})  \Big)^2, \label{eq:strictLess}
\end{align}
where the strict inequality follows from the fact that $\Re(z)< |z|$ for any non-real complex number $z$. Define a new vector $\mathbf v'=(1+\epsilon) \mathbf v^{\star} e^{-j\phi}$, with $\epsilon>0$ and arbitrarily small. The strict inequality in \eqref{eq:strictLess} implies that $\mathbf v'$ is also feasible to the relaxed problem of $\eqref{eq:imeq0}$. Furthermore, it obviously achieves a larger objective value than $\mathbf v^{\star}$. This thus contradicts the assumption that $\mathbf v^{\star}$ is the optimal solution. Therefore, the constraint $\Im (\hat{\mathbf f}^H \mathbf v)=0$ is guaranteed by the relaxed problem of \eqref{eq:imeq0} and hence it is redundant. %, which both are equivalent to problem (P2.1) as well.
 Furthermore, (P1.2) follows from a simple reformulation of the relaxed problem of \eqref{eq:imeq0}.
 % using the fact that $\hat{\mathbf f}^H\mathbf v$ is real. %, problem \eqref{P:relaxed}  can be equivalently written as (P1.4).
%Thus, Lemma~\ref{lemma:equivalance} follows. %This completes the proof of Lemma~\ref{lemma:equivalance}.
\end{IEEEproof}

With Lemma~\ref{lemma:equivalance}, finding the optimal solution to (P1.1) is tantamount to solving (P1.2), which can be reformulated as %. By reformulating the  constraint \eqref{eq:P1.4Constr}, it can be verified that problem (P1.4) can be equivalently written as
\begin{equation}\label{P:1.5}
\begin{aligned}
 \quad \underset{\mathbf v}{\max}\ & |\mathbf g^H \mathbf v|^2, \quad
\text{ s.t. }
\left \|\mathbf F^{1/2} \mathbf v - \mathbf b\right\|^2\leq \beta,
\end{aligned}
\end{equation}
where %$\mathbf F\triangleq \mathbf I -a \hat{\mathbf f}\hat{\mathbf f}^H=\mathbf I -\frac{P_s \|\mathbf h\|^2}{P_s \|\mathbf h\|^2 + \sigma_r^2}\eta \mathbf f\mathbf f^H$,
$\mathbf F\triangleq \mathbf I -a \hat{\mathbf f}\hat{\mathbf f}^H$, $\mathbf b\triangleq a \mathbf F^{-1/2} \hat{\mathbf f}$, and $\beta\triangleq a+\|\mathbf b\|^2$.
%Note that since $\eta \|\mathbf f\|^2<1$, i.e., the recycled energy must be smaller than that transmitted, $\mathbf F$ is positive definite and hence invertible. With eigenvalue decomposition, it can be obtained that
%\begin{align}
%\mathbf F^{-1} = \mathbf I + \frac{a \hat{\mathbf f}\hat{\mathbf f}^H}{1-a \|\hat {\mathbf f}\|^2}.\label{eq:Finv}
%\end{align}
By defining a new optimization vector
$\hat{\mathbf v} \triangleq (\mathbf F^{1/2}\mathbf v-\mathbf b)/\sqrt{\beta}$, problem \eqref{P:1.5} is then equivalent to
\begin{equation}\label{P:1.6}
\begin{aligned}
 \quad \underset{\hat{\mathbf v}}{\max}\ & |\mathbf g^H \mathbf F^{-1/2} (\mathbf b+\sqrt{\beta }\hat{\mathbf v})|^2, \quad
\text{ s.t. }
\left \|\hat{\mathbf v}\right\|^2\leq 1.
\end{aligned}
\end{equation}
The optimal solution to problem \eqref{P:1.6} can be obtained with the following inequalities:
\begin{align}
\big |\mathbf g^H \mathbf F^{-1/2} (\mathbf b+\sqrt{\beta }\hat{\mathbf v})\big |& \leq
\big |\mathbf g^H  \mathbf F^{-1/2} \mathbf b \big | + \sqrt{\beta}\big |\mathbf g^H \mathbf F^{-1/2}\hat{\mathbf v}\big | \notag \\ %\label{eq:triangleqIneq}\\
& \leq \big |\mathbf g^H  \mathbf F^{-1/2} \mathbf b \big | + \sqrt{\beta} \big \|\mathbf F^{-1/2} \mathbf g \big \|, \notag % \label{eq:Caucy}
\end{align}
where the upper bound is attained by setting
%where we have used the triangle and Cauchy-Schwarz inequalities in \eqref{eq:triangleqIneq} and \eqref{eq:Caucy}, respectively. It can be verified that equality holds for both \eqref{eq:triangleqIneq} and \eqref{eq:Caucy} if $\hat{\mathbf v}$ is set as
%\begin{align}
$\hat{\mathbf v}^{\star} = \mathbf F^{-1/2} \mathbf g e^{j\psi}/\big \| \mathbf F^{-1/2} \mathbf g\|$, %\label{eq:hatopt}
%\end{align}
with $\psi = \angle \mathbf g^H \mathbf F^{-1/2} \mathbf b$. %=\angle \mathbf g^H \mathbf F^{-1} \mathbf f=\angle \mathbf g^H \mathbf f$.
%Therefore, $\hat{\mathbf v}^{\star}$ in \eqref{eq:hatopt} is the optimal solution to problem \eqref{P:1.6}. %Based on \eqref{eq:b} and \eqref{eq:vTransformed},
 As a result, the optimal solution to problem \eqref{P:1.5}, and hence that to the original problem (P1.1) can be obtained as
\begin{align}
\mathbf v^{\star} &= \mathbf F^{-1/2} (\mathbf b+\sqrt{\beta }\hat{\mathbf v}^{\star})
= a \mathbf F^{-1} \hat{\mathbf f} + \frac{\sqrt{\beta}\mathbf F^{-1} \mathbf g  e^{j \psi}}{\|\mathbf F^{-1/2} \mathbf g \|}. \label{eq:vstarII}
\end{align}
By evaluating \eqref{eq:vstarII} with the identity $\mathbf F^{-1} = \mathbf I + a \hat{\mathbf f}\hat{\mathbf f}^H/(1-a \|\hat {\mathbf f}\|^2)$ and after some manipulations,
 the results given in Theorem~\ref{theo:solP1} can be obtained.

\section{Proof of Theorem~\ref{theo:solP2}}\label{A:solP2}
The second-order derivative of $R(\alpha)$ can be obtained as
\begin{align}
R''(\alpha)= \frac{-\gamma_1 C\Big(C(1-\alpha)(\gamma_1+2)+2\alpha(1+\gamma_1) \Big)}{2 \ln 2 \big(\alpha + C(1-\alpha)+\gamma_1 \alpha \big)^2 \big(\alpha+C(1-\alpha) \big)^2}<0.\notag
\end{align}
Thus, (P2) is a convex optimization problem, whose solution is given by the stationary point in the interval $(0,1)$ satisfying $R'(\alpha)=0$, or
\begin{align}
\frac{\gamma_1 C (1-\alpha)}{\Big(\alpha+C(1-\alpha)+\gamma_1 \alpha\Big)\Big(\alpha+C(1-\alpha)\Big)}=\ln\left(1+ \frac{\gamma_1 \alpha}{\alpha+ C(1-\alpha)} \right). \label{eq:stionaryPt}
\end{align}
By letting $z=1+ \frac{\gamma_1 \alpha}{\alpha+ C(1-\alpha)}$, we then have $1<z<1+\gamma_1$. Furthermore, it can be shown  that equation \eqref{eq:stionaryPt} is equivalent to $f(z)=0$, with $f(z)$ defined in \eqref{eq:fz}. The properties of $f(z)$ given  in Theorem~\ref{theo:solP2} can then be verified.
  %The corresponding throughput expression in \eqref{eq:R2star} can be obtained as well.

%This completes the proof of Theorem~\ref{theo:solP1}

\bibliographystyle{IEEEtran}
\bibliography{IEEEabrv,IEEEfull}

\end{document}